\documentclass[showpacs,prl,twocolumn,superscriptaddress]{revtex4}
\usepackage{epsfig}
\usepackage{bm}

\begin{document}

\title{Classical dimers and dimerized superstructure in  orbitally degenerate honeycomb antiferromagnet}
\author{G.~Jackeli}
\altaffiliation[]{Also at E. Andronikashvili Institute of Physics, 0177
  Tbilisi, Georgia } 
\affiliation{Max-Planck-Institut f\"ur Festk\"orperforschung,
  Heisenbergstrasse 1, D-70569 Stuttgart, Germany}
\author{D.~I.~Khomskii}

\altaffiliation[]{Also at the Loughborough University, UK}

\affiliation{II. Physikalisches Institut, Universit\"at zu K\"oln, Z\"ulpicher Strasse 77, 50937 K\"oln, Germany}

\date{\today}
\begin{abstract}
We discuss the ground state of the spin-orbital model  for  spin-one ions with partially filled
$t_{2g}$ levels on a honeycomb lattice. We find that the orbital degrees of freedom  induce a
spontaneous dimerization of spins and drive them into nonmagnetic manifold spanned by hard-core
dimer (spin-singlet) coverings of the lattice. The cooperative ``dimer Jahn-Teller'' effect is introduced through a
magnetoelastic coupling and is shown to lift the orientational degeneracy of
dimers leading to a peculiar valence bond crystal pattern. The present theory provides a
theoretical explanation of nonmagnetic dimerized superstructure experimentally seen in
Li$_2$RuO$_3$ compound at low temperatures.  
\end{abstract}
\maketitle

The nearest-neighbor Heisenberg antiferromagnet on a bipartite lattice has a
N\'eel-type  magnetically long-range ordered ground state. 
However, such a classical order of spins can be destabilized by
introducing a frustration into the system through the competing interactions
 that may lead  to the  extensively degenerate classical ground states \cite{frusrev}. 
In such systems exotic quantum phases without long-range order can
emerge as the true ground states.
In this Letter we want to point out and 
discuss another scenario, that can appear
when magnetic ions on a bipartite lattice possess also an orbital degeneracy.
The physics of such systems
may be drastically different from that of pure spin models, as the occurrence
of an orbital ordering can modulate the spin exchange and preclude the
formation of magnetically  ordered state  on a bipartite lattice.
In the following we focus on a system with 
threefold-orbitally-degenerate $S=1$ magnetic ions on a honeycomb lattice.
This model is suitable to describe $d^2$ and $d^4$ -type transition-metal
compounds with partially filled $t_{2g}$ levels, like the layered compound
Li$_2$RuO$_3$ \cite{miura}. Here the layers are formed by edge-sharing  network of RuO$_6$
and LiO$_6$ octahedra. The Ru ions make a honeycomb lattice and Li ions reside
in the centers of hexagons. These layers are well separated by the remaining Li
ions. The magnetically active Ru$^{4+}$-ions are characterized by four electrons in the threefold degenerate
$t_{2g}$-manifold coupled into a $S=1$ state.
Li$_2$RuO$_3$ undergoes a metal-to-insulator transition on cooling below 
540 K \cite{miura}.  At the transition the 
magnetic susceptibility shows a steep decrease and its low temperature value
can be considered to be due almost entirely to the Van Vleck paramagnetism.
The structural analyses have revealed the formation of dimerized superstructure
of Ru-Ru bonds in the low temperature phase.
These observations indicate that Ruthenium spin-one degrees of freedom are
mysteriously missing at low temperatures and suggest the formation of an
unusual spin-singlet dimer phase in the ground state of the system.

Here we describe the microscopic theory behind the stabilization of such a 
spin-singlet dimer state. 
We  argue that, remarkably,  such a  novel phase can be realized 
on a  honeycomb lattice  because of orbital degeneracy, without invoking
any exotic spin-only interactions.
A possibility of formation of orbitally driven  magnetically disordered states 
has been suggested within various coupled spin-orbital models
\cite{pen,feiner,Timy1,mila,my,KM, DKh}. The orbital induced frustration in $t_{2g}$
based systems on a  bipartite (cubic) lattice has also been considered
\cite{K1,K2}. The emergence of new phases due to the $p$-orbital degeneracy
of cold atoms in optical lattices has been recently discussed within a spinless
fermion model on a honeycomb and other two-dimensional lattices
\cite{porb1,porb2}. Yet, the peculiar case of  partially filled
$t_{2g}$ levels on a honeycomb lattice leads to new results: 
the onset of an orbitally driven spin-singlet dimer phase in a spin-one system.

{\it The model}.-- We assume that the low-temperature insulating 
phase of Li$_2$RuO$_3$ is of Mott-Hubbard type and describe the low energy physics 
within the Kugel-Khomskii type spin-orbital Hamiltonian \cite{KK}. We consider undistorted
honeycomb lattice of Ru ions and look for possible instabilities towards symmetry
reductions. In the  Li$_2$RuO$_3$ crystal structure three distinct bonds of
honeycomb lattice are in $xy$, $xz$, and $yz$ planes (in cubic
notations). We consider the leading part of the nearest-neighbor (NN) hopping
integral of  $t_{2g}$ orbitals ($d_{xy}$,
$d_{yz}$, and $d_{xz}$) due to the direct $\sigma$-type overlap.
The $dd\sigma$ overlap in $\alpha\beta$ plane
($\alpha,~\beta=x,~y,~z$) connects only the orbitals of same $\alpha\beta$
type. 
The effective spin-orbital Hamiltonian for such a system 
been reported in Refs.~\cite{motome,Vmy}. It has the following form:
\begin{eqnarray}
H=\sum_{\langle ij\rangle}{\Big \{} J{\big [}\vec S_i\cdot \vec S_{j}-1{\big ]}{\bar O}_{ij}
-{\big [}J_0 \vec S_i\cdot \vec S_{j}+J_1{\big ]} O_{ij}
{\Big \}},
\label{eq1}
\end{eqnarray}
where the sum is taken over pairs of NN sites, $\vec S_i$
are spin-one operators, and the orbital contribution are described by ${\bar O}_{ij}$
and $O_{ij}$ operators.
The second-order virtual processes locally conserve orbital index.
The orbital degrees are thus static Potts-like variables and their contribution can be expressed simply
in terms of projectors $P_{i,\alpha\beta}$ onto the singly occupied orbital state $\alpha\beta$
at site $i$. With this definition of the  projectors the orbital part
of the Hamiltonian can be written in  the form  equally valid for $t_{2g}^2$ and its
particle-hole symmetry related $t_{2g}^4$ configurations.
 The orbital operators ${\bar O}_{ij}$ and $O_{ij}$ 
along the bond $ij$ in $\alpha\beta$-plane are given by:
$\bar{O}_{ij}=P_{i,\alpha\beta}P_{j,\alpha\beta}$ and $O_{ij}= P_{i,\alpha\beta}(1-P_{j,\alpha\beta})+
P_{j,\alpha\beta}(1-P_{i,\alpha\beta})$.  
For further analysis it is convenient to rewrite the Hamiltonian as the sum of three terms:
$H=E_0+H_{\rm AF} + H_{\rm FM}$,
\begin{eqnarray}
H_{\rm AF}=\!\!\sum_{\langle ij\rangle}J{\big [} \vec S_i\cdot \vec S_{j}+
\zeta   {\big ]}{\bar O}_{ij}\, ,
H_{\rm FM}=\!\! - J_{0} \sum_{\langle ij\rangle} \vec S_i\cdot \vec
S_{j}O_{ij}\, ,
\nonumber\\
\label{eq2}
\end{eqnarray}
$E_{0}=-2J_{1}N$, $N$ is number of lattice sites, and  $\zeta=2 J_1/J-1$ \cite{note1}.  
For the full expressions
of the exchange constants in terms of hopping integral $t$, on-site  Coulomb repulsion
$U$ and  Hund's coupling $J_H$ we refer reader to
Refs.~\cite{motome,Vmy}.
To the leading order in small parameter $\eta=J_{\rm H}/U$,
the coupling constants are given by $J\approx (1-\eta)\frac{t^2}{U}$, 
$J_0\approx \eta \frac{t^2}{U} $, $J_1\approx (1+2\eta)\frac{t^2}{U}$, and
$\zeta \approx 1+6\eta$.  
For our further analysis, it will be important that $\zeta\ge 1$ for all values
of the Hund's coupling. The inequality implies that any magnetically ordered
ground state has a positive classical energy on antiferromagnetic (AF) bonds.
It suggests that the formation of a low dimensional network of AF
spin-coupling patterns are energetically favorable due to a larger
gain of quantum spin-energy per bond. This picture is substantiated by the
exact solution of the ground state problem at zero Hund's coupling described
in the next section.

{\it Zero Hund's coupling.}-- 
The spin-orbital model defined in Eq.~(\ref{eq2})
has a small parameter $\eta \ll 1$. We start our analysis from the limit of
zero Hund's coupling  $\eta=0$ ($H_{\rm FM}$=0) and look for the possible
spin-coupling patterns generated by $H_{\rm AF}$ term of the Hamiltonian.  

\begin{figure}
\epsfysize=40mm
\centerline{\epsffile{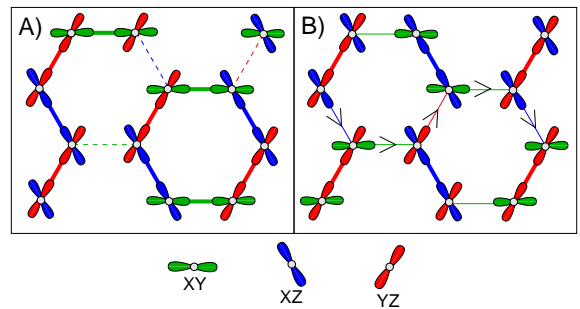}}
\caption{(Color online) 
Examples of orbital and spin-coupling patterns on the
honeycomb lattice of Ru ions. A) Decoupled AF chain and ring
with corresponding orbital pattern.
B) An example of the spin-singlet dimer covering minimizing the energy at zero Hund's coupling.
Thick (thin) lines denote  AF (FM) intra- (inter-)dimer bonds, respectively.
Dashed lines stand for the noninteracting bonds.}
\label{fig:oo}
\end{figure}

The antiferromagnetic  term $H_{\rm AF}$ is active only on the bonds with
corresponding orbital being singly occupied at both ends of a bond ({\it
  e.g} on a bond $ij$
in $xy$ plane both sites  have $xy$ orbital singly occupied).
The AF bonds can only form non-intersecting linear open and/or closed chains (see Fig.~\ref{fig:oo}A).
The longer is the chain, the more energy can be gained from the AF spin
interaction. However, for each AF bond we pay a positive energy $\zeta \ge 1$
(the second term in $H_{\rm AF}$). Under these conditions the minimal possible AF energy is achieved when all
AF chains are dimers. The  proof of the stability of dimer states against the
formation of longer open Heisenberg chains is based on  the variational estimate on the ground-state energy
of the $M$-site spin-one AF Heisenberg chain with open ends:
$E_M \ge 1-\frac{3M}{2}$, with the equality attained
only at $M=2$. This estimate can be obtained by dividing the chain into shorter overlapping
sub-chains of lengths two and three with exactly known energies $E_2=-2$ and
$E_3=-3$. The dimer states have a lower energy than the  closed chains
if ${\bar E}_M>-\frac{3M}{2}$, where ${\bar E}_M$ is the ground-state energy
of the $M$-site spin-one AF Heisenberg ring. The last inequality is satisfied for any 
AF Heisenberg ring with $M>4$ (see for example Ref.~\cite{neirotti}). Note that on a honeycomb lattice only the
closed chains with $M\ge 6$ can be formed.
Therefore, at zero Hund's coupling  $\eta=0$, in the
ground state manifold the  pattern of AF bonds corresponds
to a {\it hard-core dimer covering} of the lattice and 
{\it on each inter-dimer bond  only one site  
has  active orbital  singly occupied}.  The  $H_{\rm AF}$ term  is inactive on such inter-dimer bonds, 
and spins of different dimers are decoupled. Spins are coupled into quantum spin-singlet state on dimer
bonds and are thus gapped. Such a spin-singlet dimer states form the exact ground
state manifold of spin-orbital Hamiltonian  Eq.~(\ref{eq2}) in the limit of zero Hund's
coupling. 

{\it Ground state manifold.}--
The ground state manifold is extensively degenerate: there are infinitely
many ways of covering a honeycomb lattice with hard-core dimers, and each dimer covering
has its own  Ising-type degeneracy connected with the orientation of an ``inactive'' orbital.
Thus, in contrast to the case of one d-electron (S=1/2) where the state is
determined by the covering of the lattice with  spin-singlet dimers \cite{my},
here we have an extra degeneracy. At each site the second orbital points along
one of two remaining bonds 
(the arrows in Fig.1B), with the rule that at each of such ``empty''
 bonds there should be only one orbital (one arrow).
For any  dimer covering  the inter-dimer (``empty'') bonds form non-intersecting
linear chains.  Along an inter-dimer path all arrows must be directing along the same
direction [see as an example the zig-zag like
linear chain in Fig.~\ref{fig:oo}B].
However, on each inter-dimer path we can  inverse all the arrows
and still remain in the ground state manifold. Thus each inter-dimer path has
a two-fold degeneracy of Ising-type.
The degeneracy of a state with given dimer covering is thus equal to 
$2^{N_c}$, where $N_{c}$ is a  number of decoupled inter-dimer chains.
One finds that $N_{c}\sim \sqrt{N}$ for all dimer coverings except one case when 
all inter-dimer chains are hexagonal loops. In this case the honeycomb lattice
is divided into non-overlapping hexagons with no dimers, and  the remaining
hexagons are occupied by three dimers. This gives $N_{c}=N/6$ leading to a
finite contribution to a bulk entropy from  Ising-type degeneracy.
The above degeneracy is, however, easily lifted by any interaction which induces
a non-zero coupling of non-active orbitals on dimer bonds and thus correlates the
 Ising-like variables (direction of arrows) of neighboring inter-dimer chains.
For example, the coupling of Jahn-Teller distortions on NN edge sharing oxygen 
octahedra favors the ferro-type orbital order of singly occupied 
non-active orbitals on dimer bonds (see Fig.~\ref{fig:oo}B) \cite{note2}. 
The $\pi$-type contribution to the hopping integral instead stabilizes an antiferro-type  orbital order
by increasing the AF coupling on a dimer bond \cite{note3}. 
In real materials the two mechanisms will have  different energy scales
and the dominant one will dictate the resulting orbital pattern.
In both cases one completely lifts the degeneracy of a given dimer state.

The degeneracy of second type originates from the orientational degeneracy of dimers
and is exactly equal to the number of hard-core dimer coverings of the honeycomb
lattice.  The latter is well known to be an extensive quantity.
The classical dimers on a honeycomb lattice exhibit  a power-law decay of
dimer-dimer correlation function and are thus in a critical state
\cite{moessner}.
Therefore, one expects that any perturbation which may favor  one or another type of dimer
orientation  may  lift this extensive degeneracy, resulting in a long-range
ordered pattern of dimers. We have considered a possibility of perturbing 
the exact dimer ground state manifold by small but finite Hund's coupling
$0<\eta\ll 1$. The Hund's coupling produces a ferromagnetic (FM) term $H_{\rm FM}$
active on inter-dimer bonds (see Fig.~\ref{fig:oo}B). At small $\eta$, the dimer state is stable against  
the weak FM inter-dimer interaction.  In this case the magnetic  
contribution along the FM bond is zero ($\langle \vec{S}_i\cdot \vec{S}_{j}\rangle=0$
for $i$ and $j$ belonging to different dimers). However, the weak inter-dimer coupling
may, in principle, lift the orientational degeneracy through order out of disorder by triplet
fluctuations \cite{my}. We find that on a honeycomb lattice these quantum
fluctuations do not fully lift the dimer degeneracy. They only disfavor the
configurations with three dimers on a hexagon because such a hexagonal loop
of alternating AF and FM bonds is frustrated in the classical limit.
In the next section we introduce a magnetoelastic coupling and show that it
fully lifts the orientational degeneracy of dimers. 

{\it Degeneracy breaking  by magnetoelastic coupling}.-- 
The magnetoelastic mechanism of lifting the extensive degeneracy of
frustrated systems has been successful in explaining the experimentally observed
structures  \cite{Timy1,Oleg,Penc,Balents}. 
The physics behind this mechanism is
that the correlated nature of structural distortions that appear due to the
modulation of magnetic energies on bonds may select a particular pattern
of distortions and thus lift the extensive degeneracy of the ground state manifold.

\begin{figure}
\epsfysize=50mm
\centerline{\epsffile{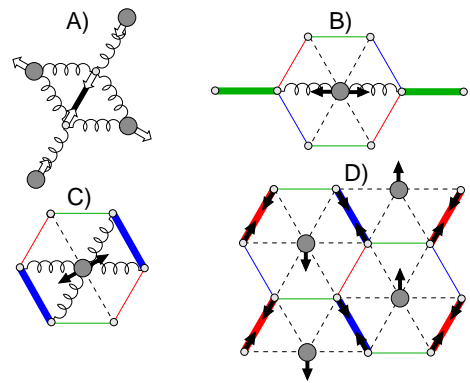}}
\caption{(Color online) Lifting dimer degeneracy by magnetoelastic coupling.
The small light (large dark) circles denote Ru (Li) ions. The arrows show the
displacements of corresponding ions. A) The sketch of Li displacements induced  by dimerized
  Ru-Ru bond. B) and C) Two examples of destructive interference of Li displacements 
 induced by neighboring dimers. D) The ground state dimer pattern selected by
 magnetoelastic coupling. This pattern exactly corresponds to the one found in Li$_2$RuO$_3$ in Ref.\cite{miura}.}
\label{fig:dist}
\end{figure}
In the ground state manifold the spin degrees of freedom are described by the
product of spin-singlet dimer states. 
A  shortening 
of  a bond where the
singlet is located enlarges the  magnetic energy gain, because of the increase in the
exchange coupling $J$ on that bond. The magnetic energy gain, being linear in distortion, 
outweights the increase in elastic energy and 
would always lead to such a contraction of singlet bonds for any dimer
covering. But due to elastic coupling of these distortions different distorted patterns will 
lead to different elastic energy and hence to the lifting of dimer degeneracy. We, first, formalize this
picture  within the simplest model and later discuss its extension. 
The schematic structure of Li$_2$RuO$_3$ is shown in Fig.2D.
In the minimal model  
the exchange coupling is assumed to depend solely on a distance between Ru
ions, and the lattice degrees of freedom are described  within Einstein 
phonon model for Ru ions:
\begin{eqnarray}
{\cal E}_{\rm ME}=-\gamma\epsilon\sum_{\langle ij\rangle}{\big [}\vec u_i- \vec
v_{j}{\big ]}\vec l_{ij}+\frac{k}{2}\sum_{i}{\vec u}_{i}^{2}
\label{ee3}
\end{eqnarray}
where $\gamma=-\frac{ 1}{J}\frac{\partial J(r)}{\partial r}$ is a magnetoelastic
coupling constant, $\epsilon$ is magnetic energy gain on a dimer bond,
${\vec u}_{i}$ is a displacement of Ru ions at site $i$, and  $k$ is Einstein
phonon constant. The dimer positions are described by  $\vec l_{ij}$: the latter 
is an unit vector along a bond $ij$ occupied by dimer and is zero
otherwise. We minimize the  energy
functional ${\cal E}_{\rm ME}$ with respect to set of ${\vec u}_{i}$ and
eliminate them in favor of $\vec l_{ij}$. We find  
${\cal E}_{\rm ME}=-\frac{k}{2}\sum_{i}{\vec {\bar u}}_{i}^{2}$ ,
  and ${\vec {\bar u}}_{i}=\frac{\gamma\epsilon}{k}\sum_{j}\vec l_{ij}$.
The set of  ${\vec {\bar u}}_{i}$ define the distortion pattern in which Ru
ions on dimer bonds are moved towards each other. However, as the dimer bonds
do not share a common site, ${\cal E}_{\rm ME}$ is a constant. It is
independent of dimer variables $\vec l_{ij}$. In order to introduce the
coupling between the dimers we extend the model by including a finite force
induced on Li ions by distorted pattern of Ru ions.     
Consider an isolated complex of a Ru-Ru bond together with neighboring Li ions, shown in Fig.~\ref{fig:dist}A.
When this bond is occupied by spin-singlet dimer it  gets  contracted, and the
displacements of Ru ions  induce the corresponding distortion 
of Li complex as seen in Fig.~\ref{fig:dist}A.
On a lattice  the distorted dimer bonds 
will now be coupled through the force induced on a common Li ion, an effect
similar to a cooperative Jahn-Teller physics.
The dimer configurations for which the induced forces on Li sites interfere
in a non-destructive manner result in a larger distortion and hence more
gain in energy. The two orientations of neighboring dimers shown in
Fig.~\ref{fig:dist}B and C are energetically unfavorable, as in both cases the   
induced forces exactly cancel each other. Thus the ground state 
dimer pattern should satisfy the constraint of no-$B$ and no-$C$ type
configurations. Note that a dimer covering satisfying 
the former constraint automatically satisfies the latter. 
It is possible to check that the only 
 possibility to  fulfill such a  no-$B$ constraint is realized for the dimer pattern
shown  in Fig.~\ref{fig:dist}D. This dimerization pattern exactly reproduces the one
observed in the insulating phase of Li$_2$RuO$_3$ \cite{miura}.

We note that above described mechanism of the selection of spin-singlet dimer
pattern equally applies to the case when singlets are formed by spin-one-half
degrees of freedom. The spin-singlet dimer nature of the ground state
manifold for $d^1$ systems on a honeycomb lattice has been proven in
Ref.~\cite{my}. We therefore predict the same dimerized
superstructure also for  $d^1$ systems, such as V$^{4+}$ and Ti$^{3+}$ based
compounds with such structure (if orbital degeneracy will not be lifted by some external mechanism like trigonal distortion, often present in such structure). 


{\it Summary}.-- To summarize, we have studied the ground state of spin-one honeycomb
antiferromagnet with partially filled $t_{2g}$ levels. We have demonstrated
that the orbital degeneracy induces spontaneous dimerization of spins and
drives them into extensively degenerate manifold of spin-singlet dimer states.
The orientational degeneracy of dimers is then lifted through the
magnetoelastic interaction that stabilizes a peculiar valence bond crystal
state. Our theory provides an explanation for the observed 
nonmagnetic dimerized superstructure in Li$_2$RuO$_3$ compound.

We are grateful to G. Khaliullin and H. Takagi for useful discussions.
We acknowledge kind hospitality at KITP, UCSB where the part of this work
has been done. This research was supported in part by the National Science  
Foundation under Grant No. PHY05-51164.
G.J. acknowledges support by GNSF under the Grant No.06-81-4-100; the work of D.Kh. 
is supported by SFB 608 and by the European project COMEPHS.



\begin{thebibliography}{20}
\bibitem{note1} In deriving Eq.~(\ref{eq2}) 
$\sum_{\langle ij\rangle}[2{\bar O}_{ij}+O_{ij}]=2N$ was used.
\end{thebibliography}
\end{document}